\documentclass{article}[12pt]

\input{epsf}
\usepackage{epsf}
\usepackage{graphicx,epsfig}
\usepackage{bm}
\usepackage{latexsym,float,amsmath,amssymb}

\newcommand {\ga} {\ {\raise-.5ex\hbox{$\buildrel>\over\sim$}}\ }
\newcommand {\la} {\ {\raise-.5ex\hbox{$\buildrel<\over\sim$}}\ } 

\newcommand{\fig}[1] {Fig.~(\ref{#1})}

\def\be{\begin{equation}}
\def\ee{\end{equation}}
\def\ba{\begin{eqnarray}}
\def\ea{\end{eqnarray}}
\renewcommand{\(}{\left(} 
\renewcommand{\)}{\right)}

\renewcommand{\Psi}{\varPsi}

\begin{document}

\thispagestyle{empty}

\begin{center}

{\Large \bf Dark Energy, with Signatures}

\bigskip
\bigskip
\bigskip

{\large \sc Sourish Dutta\footnote{\tt sourish.d@gmail.com} and Robert~J.~Scherrer\footnote{\tt robert.scherrer@vanderbilt.edu}}

{\it Department of Physics and Astronomy, Vanderbilt University, Nashville,
TN-37235}

{\large \sc Stephen~D.~H.~Hsu\footnote{\tt hsu@uoregon.edu}} 

{\it Institute of Theoretical Science, University of Oregon,
Eugene, OR 97403-5203}

\vspace*{1 cm}

{\small Essay written for the Gravity Research Foundation 2010 Awards for Essays on Gravitation}\\
{\small Submitted on \today}\\

\vspace*{1 cm}

\large{\bf Abstract}
\end{center}
\noindent
We propose a class of simple dark energy models which 
predict a late-time dark radiation component and 
a distinctive time-dependent equation of state $w(z)$ for redshift $z < 3$.
The dark energy field can be coupled strongly enough to
Standard Model particles to be detected in colliders, and
the model requires only modest additional particle content
and little or no fine-tuning other than a new energy scale
of order milli-electron volts.

\newpage

\setcounter{page}{1}

\noindent
\section*{Introduction}
A slew of recent cosmological observations over the past decade have led to a spectacular conclusion: the energy density of our Universe seems to be dominated by an unknown negative-pressure ``dark energy'' which is accelerating its expansion.


The simplest form of dark energy is of course Einstein's cosmological constant. However, this explanation has severe fine-tuning problems, compelling a search for other possibilities. The fact that the observed vacuum energy happens to be just a few times greater than the present matter density has led to speculations that it might in fact be evolving with time -- only now reaching a value comparable to the matter density. Such a time-varying dark energy is referred to as \emph{quintessence}. The simplest realization of this scenario is through the use of slowly rolling canonical scalar fields \cite{RatraPeebles,TurnerWhite,CaldwellDaveSteinhardt,LiddleScherrer,SteinhardtWangZlatev,Dutta1}.

In most quintessence models, however, the field(s) responsible for the acceleration has to be almost completely decoupled from the rest of the Universe. This is disappointing, since it suggests that direct detection of quintessence through its interactions with Standard Model particles will be extremely challenging, perhaps impossible.

In this essay we discuss a quintessence scenario in which the dark energy field
can be coupled strongly enough to Standard Model particles to be detected in
colliders, and which allows for a significant time variation in the dark energy
equation of state parameter $w$ \cite{DuttaHsuReebScherrer}. The time-variation
in $w$ has a characteristic form which depends on only a single parameter, and
can thus be excluded or confirmed by cosmological observations in the near future. Our model only requires a singlet scalar field (or, among other possibilities, a small gauge sector like $SU(3)$ Yang-Mills) and a new energy scale on the order of milli-electron volts. 

\section*{The Model}
\label{aa}
\begin{figure}
\begin{center}
	\epsfig{file=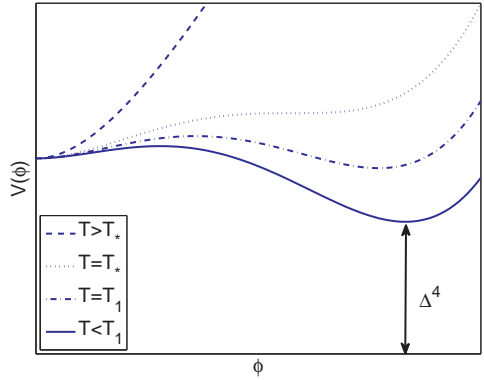,height=55mm}
	\caption{\label{potential}\textit{An example for the evolution of the finite temperature effective potential $V\(\phi,T\)$ of the dark energy field, eqn.~(\ref{effpot}), as the temperature $T$ decreases through the first order phase transition region $\sim T_1$.}}
	\end{center}
\end{figure}

The simplest realization of our model is through the use of a singlet scalar field $\phi$, which is allowed to be non-trivially coupled to Standard Model particles, at least compared to a quintessence field which can only have Planck-suppressed couplings. The finite temperature effective potential, which includes interactions of this field with virtual particles and the heat bath, can be taken to be similar to the Higgs potential in the electroweak phase transition (see \fig{potential}):

\begin{equation}
V \(\phi, T\) =A+D\(T^2-T_{2}^2\)\phi^2  - E T \phi^3 + \frac{1}{4}\lambda
\phi^4~,\label{effpot}
\end{equation}
where $D$, $E$, $\lambda$ and $A$ are constants. $A$ can be adjusted to give the correct value of the observed dark energy density when $T=0$.
$T_2$ is defined as the temperature where $V''\(\phi=0\)=0$. We choose $T_2$ to correspond to the energy scale of the cosmological constant ($\Delta$), i.e.~$T_2~\sim11.6\,\text{K}$, and assume that it represents a new energy scale in particle physics. At high temperatures, $T\gg T_2$, $\phi=0$ is the only minimum of the potential. As the Universe cools down, an inflection point appears in the potential at temperature $T_*=T_2/\sqrt{1-9 E^2/8\lambda D}$. At lower temperatures, this splits into a barrier and a second minimum. The critical temperature $T_1=T_2/\sqrt{1- E^2/\lambda D}$  corresponds to the point where the second minimum is equal in (free) energy to the $\phi=0$ minimum. At temperatures $T<T_1$, the second minimum has lower free energy than the one at $\phi=0$. 

At temperatures $T> T_1$ (in terms of the redshift $z$, roughly at $z>3$) the
dark energy field remains trapped at the $\phi=0$ minimum, providing a constant
energy density, which we assume to be slightly higher than $\Delta^4$. As the
temperature approaches $T_1$ and below, a first order phase transition is
triggered as bubbles of the new phase are nucleated and expand. The physics of
the phase transition is almost identical to that of the Higgs sector in models
of electroweak baryogenesis.  This transition releases energy in relativistic
modes (i.e., scalar particles of the $\phi$ field), and brings the vacuum energy
to $\Delta$. Because the correlation length of the transition is microscopic,
and the relativistic modes couple weakly to ordinary matter (i.e., more weakly
than photons, perhaps similar to neutrinos), such a transition is only loosely
constrained by observation. The positive pressure of the radiation, which
eventually redshifts away, causes the effective $w$ of the dark energy to vary
in redshift (or equivalently, in time).

The most important feature of our model is that it has a weakly first order phase transition at a temperature of order $\Delta$, which is natural if one assumes the dynamics of $\phi$ to be entirely determined by that energy scale and dimensionless couplings of order one. It is an interesting coincidence that this occurs at a redshift of $z \sim 3$ if the temperature of the dark energy field is similar to that of the Standard Model particles. This need not be the case, but it seems a reasonable assumption, especially if there are non-negligible interactions between $\phi$ and ordinary particles, which would enforce thermal equilibrium at sufficiently high temperatures. When the transition happens at $z \sim 3$ the resulting radiation component leads to significant and characteristic variation in $w(z)$. The form of $w(z)$ is determined by a single parameter $f$ -- the energy fraction in relativistic dark radiation modes just after the phase transition. In some cases, such as the gauge models discussed below, even this fraction is calculable.

Any sector which produces a weakly first order transition at a temperature of order $\Delta$  would also suffice. For example, pure $SU(N)$ gauge theories with $N > 2$ have first order deconfinement phase transitions \cite{gauge} and exhibit effective potentials like those in Fig.~(\ref{potential}), with $\phi$ an order parameter for confinement, for example the Polyakov loop. Here, the latent heat and fraction of energy in relativistic modes is calculable via lattice simulation.

Note that our model does not in any way {\it explain} the existence of the energy scale $\Delta$, or why it determines the vacuum energy density today. In particular, why should the vacuum energies from all the other degrees of freedom cancel out, leaving the dark energy field to determine the cosmological constant? One way of explaining this would be to assume that somewhere in the configuration space, outside the region depicted in \fig{potential}, the potential reaches a global minimum $V\(\phi_* \)=0$, where as a result of some unknown mechanism the total vacuum energy (including zero point energies and radiative corrections from {\it all} fields) is exactly zero, leaving  of $V \( \phi \)$ as the only vacuum energy. In the string theory Landscape, which exhibits many vacua and energy splittings smaller than $\Delta$, this scenario is quite natural.

In any case, if one assumes that new physics at the energy scale $\Delta$ determines the observed cosmological constant, it is easy to obtain a predictable redshift-dependent $w=w(z)$ together with interesting particle physics signatures -- no fine tuning of parameters is required. 

\section*{Observational Consequences}
\label{Observational Consequences}

\begin{figure}
\begin{center}
	\epsfig{file=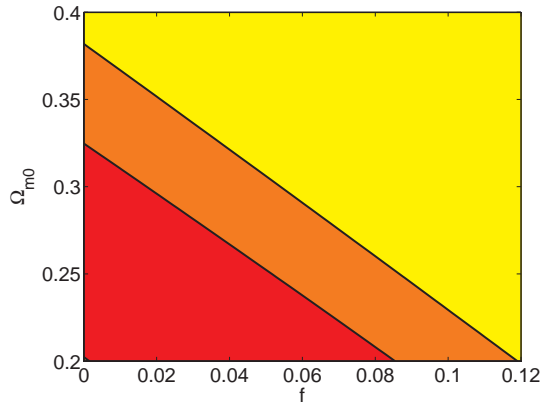,height=55mm}
	\caption
	{\label{likelihood}\textit{Likelihood contour for the parameters $f$ and $\Omega_{m0}$ (the matter density fraction of the Universe) based on data from Type Ia Supernovae \cite{Davis}. The yellow (light) region
is excluded at the 2$\sigma$ level and the orange (darker) region
is excluded at the 1$\sigma$ level.  The red (darkest) region is
not excluded at either confidence level. Clearly, the data do not rule out a sizable fraction of the dark energy today being in relativistic modes.}}
\end{center}
\end{figure}

\paragraph{Astrophysics}
As discussed in the previous section, our model produces a certain amount of dark radiation at redshift $z\sim 3$. In this regard, our model superficially resembles other models with a dark radiation component,
such as models with extra relativistic degrees of freedom, the
Randall-Sundrum model with dark radiation, or Ho\v{r}ava-Lifshitz cosmologies \cite{DuttaSaridakis}.  The difference, of
course, is that in our model the dark radiation arises very late, and so is not
subject to the well-known limits from Big Bang nucleosynthesis or the cosmic
microwave background.  It was noted by Zentner and Walker \cite{ZW} that if one
considers only late-time constraints on extra relativistic degrees of freedom
from SNIa data, the limits are surprisingly weak.  As shown in \fig{likelihood}, this is confirmed even with the addition of more recent SNIa data.

\begin{figure}
\begin{center}
	\epsfig{file=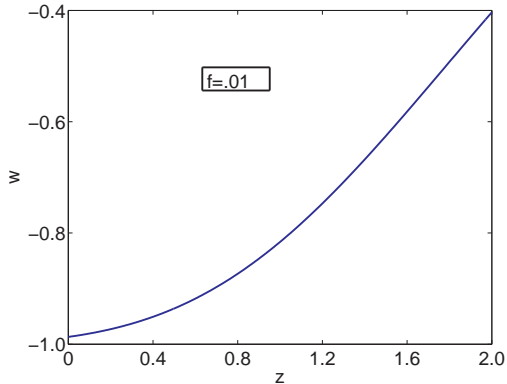,height=55mm}
	\caption
	{	\label{wz} \textit{The characteristic redshift dependence of $w$ in our model, for a conservative choice of parameters ($\Omega_{m0}=0.28,f=0.01$), not excluded at the $1\sigma$ level by the SNIa data (see \fig{likelihood}).}}
\end{center}
\end{figure}

As an example of the possible strong late-time variation of $w(z)$ predicted by this model, in \fig{wz} we plot $w$ vs.~$z$ for a conservative choice of parameters ($\Omega_{m0}=0.28,f=0.01$), which is not excluded at $1\sigma$ by the SNIa data. The pressure of the relativistic component increases the effective $w$ of the dark energy component with increasing redshift. Note that in our model, the shape of the $w$ vs.~$z$ curve is fixed once $f$ is fixed.

%

\bigskip
\paragraph{Particle Physics}
An interesting feature of our scenario is that the dark energy field can be coupled relatively strongly to Standard Model particles. This makes it possible, in principle, for this kind of dark energy to be detected in colliders. 

The simplest model we considered, comprised of a singlet scalar $\phi$, has some challenges, as a direct coupling between $\phi$ and the Higgs boson operator $H^\dagger H$ cannot be excluded. This would lead to significant radiative corrections to the $\phi$ potential parameters, making the model somewhat unnatural. However, if this fine tuning is ignored, the $\phi$--$H$ coupling would provide for direct production of $\phi$ particles at colliders.

\begin{figure}
\begin{center}
	\epsfig{file=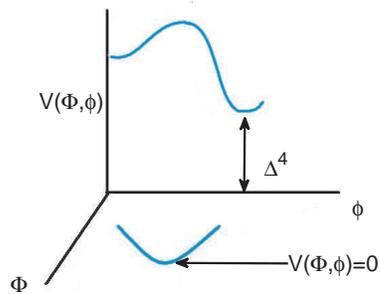,height=55mm}
	\caption
	{	\label{3dpotential} \textit{Potential energy surface for a gauge
	theory, where $\phi$ is an order parameter for confinement (Polyakov
	loop) and $\Phi$ a colored scalar field. $\Phi$ has been added to ensure
	a point in the configuration space with zero vacuum energy, but is
	otherwise unnecessary for the model. For $N > 2$, $SU(N)$ models will
	have a first order confinement-deconfinement transition as the
	temperature is lowered. However, the zero temperature deconfined phase ($\Phi = \phi = 0$) is not necessarily metastable.}}
	\end{center}
\end{figure}

An alternative possibility is to use a pure $SU(N)$ gauge theory sector ($N>2$) with strong coupling scale $\Lambda \sim \Delta$ (see \fig{3dpotential}). This model requires no fine tuning and the fraction of energy in relativistic modes after the phase transition can in principle be calculated from simulations of the $SU(N)$ theory. Glueballs of this sector would be light excitations with mass of order $\Delta$; the phase transition temperature would be at least a few times the glueball mass. The glueballs could couple to Standard Model particles via higher dimension operators. 

%

\section*{Conclusion}
\label{Conclusion}

We have discussed a class of dark energy models which have interesting cosmological as well as collider signatures. In these models, a first-order phase transition at redshift $z\sim3$ releases energy in relativistic modes (dark radiation) leading to a characteristic time-dependence in the effective dark energy equation of state. Such models are consistent with SNIa data, and are relatively easy to construct as extensions to the Standard Model.



\end{document}